\documentclass{aa}  

\usepackage{graphicx}
\usepackage{txfonts}
\usepackage{natbib}
\usepackage{multirow}

\DeclareRobustCommand{\VAN}[3]{#2}
\let\VANthebibliography\thebibliography
\def\thebibliography{\DeclareRobustCommand{\VAN}[3]{##3}\VANthebibliography}

\newcommand{\cow}{AT2018cow}
\begin{document}

   \title{A study on late time UV-emission in core collapse supernovae\\ and the implications for the peculiar transient AT2018cow}
   % \title{Comparing late time UV emission from core collapse supernovae with AT2018cow}
   \titlerunning{A study on the presence of late time UV-emission in core collapse supernovae}

   %\subtitle{I. Overviewing the $\kappa$-mechanism}

   \author{Anne Inkenhaag
          \inst{1, 2}
          \and
          Peter G. Jonker\inst{1, 2}
          \and 
          Andrew J. Levan\inst{1, 3}%\fnmsep\thanks{Just to show the usage of the elements in the author field}
          \and
          Morgan Fraser\inst{4}
          \and
          Joseph D. Lyman\inst{3}
          \and
          Llu\'is Galbany\inst{5,6}
          \and 
          Hanindyo Kuncarayakti\inst{7,8}
          }

   \institute{Department of Astrophysics/IMAPP, Radboud University Nijmegen, P.O.~Box 9010, 6500 GL Nijmegen, The Netherlands \\
              \email{a.inkenhaag@astro.ru.nl}
         \and
             SRON, Netherlands Institute for Space Research, Niels Bohrweg 4, 2333~CA, Leiden, The Netherlands
        \and
            Department of Physics, University of Warwick, Gibbet Hill Road, Coventry, CV4 7AL, UK
        \and
            School of Physics, University College Dublin, L.M.I. Main Building, Beech Hill Road, Dublin 4, D04 P7W1, Ireland
        % \and
        %      Department of Physics, University of Warwick, Coventry, CV4 7AL, UK
        \and
            Institute of Space Sciences (ICE-CSIC), Campus UAB, Carrer de Can Magrans, s/n, E-08193 Barcelona, Spain
        \and
            Institut d'Estudis Espacials de Catalunya (IEEC), E-08034 Barcelona, Spain
        \and
            Tuorla Observatory, Department of Physics and Astronomy, 20014 University of Turku, Vesilinnantie 5, Turku, Finland
        \and
            Finnish Centre for Astronomy with ESO (FINCA), 20014 University of Turku, Vesilinnantie 5, Turku, Finland
             }

   \date{Received February 7, 2024; accepted November 11, 2024}

% \abstract{}{}{}{}{} 
% 5 {} token are mandatory
 
  \abstract %might need to reduce this to 250 words
   {Over time, core-collapse supernova (CCSN) spectra become redder due to dust formation and cooling of the SN ejecta. An ultraviolet (UV) detection of a CCSN at late times thus indicates an additional physical process such as interaction between the SN ejecta and the circumstellar material, or viewing down to the central engine of the explosion. Both these models have been proposed to explain the peculiar transient AT2018cow, a luminous fast blue optical transient that has been detected in the UV 2--4 years after the event with only marginal fading over this time period.}
   {To identify if the late-time UV detection of AT2018cow could indicate that it is a CCSN, we investigate if CCSNe are detected in the UV between 2--5 years after the explosion. We determine how common late-time UV emission in CCSNe is and compare those CCSNe detected in the UV to the peculiar transient AT2018cow.} 
   {We use a sample of 51 nearby (z<0.065) CCSNe observed with the \textit{Hubble Space Telescope} within 2--5 years of discovery. We measure their brightness, or determine an upper limit on the emission through an artificial star experiment if there is no detection.}
   {For two CCSNe we detect a point source within the uncertainty region of the SN position. Both have a low chance alignment probability with bright objects within their host galaxies and are therefore likely related to the SNe, both of which are known to be interacting supernovae.
   }
   {Comparing the absolute UV magnitude of AT2018cow at late times to the absolute UV magnitudes of the two potential SN detections, there is no evidence that a late-time UV detection of AT2018cow is atypical for interacting SNe. However, when limiting the sample to CCSNe closer than AT2018cow, we see that it is brighter than the upper limits on most CCSN non-detections. Combined with a very small late time photospheric radius of \cow, this leads us to conclude that AT2018cow's late-time UV detection was not driven by interaction. It suggests instead that we are possibly viewing the inner region of the explosion, perhaps due to the long-lived presence of an accretion disc. Such properties are naturally expected in tidal disruption models and are less straightforward (though not impossible) in supernova scenarios. 
   }
   

   \maketitle
%
%-------------------------------------------------------------------

\section{Introduction}

Supernovae (SNe) are the explosive endpoints of the lives of stars. With a steep increase in the number of telescopes built specifically to detect transients (e.g., the Zwicky Transient Facility [ZTF]; \citealt{Bellm2019}, the All-Sky Automated Survey for Supernovae [ASAS-SN]; \citealt{Shappee2014}, the Asteroid Terrestrial-Impact Last Alert System [ATLAS]; \citealt{Tonry2011}, the Gravitational-wave Optical Transient Observer [GOTO]; \citealt{Gompertz2020, Steeghs2022}), the number of detected SNe has increased significantly. Over the next decade the increase in sample size will accelerate with the commissioning of the Black hole Gravitational-wave ElectroMagnetic counterpart array (BlackGEM; \citealt{Bloemen2016}) and the Vera Rubin Observatory \citep{Ivezic2019}. 

There are multiple types of SNe, observationally classified predominantly by the absence or presence of hydrogen in the explosion spectra, Type I and Type II SNe, respectively \citep{Minkowski1941}. Within Type I SNe there are multiple sub-types, corresponding to different explosion mechanisms and progenitors. Type Ia SNe are caused by a thermonuclear explosion of a white dwarf (\citealt{Nomoto1984, Woosley&Weaver1986}, and references therein). Their explosion spectra show strong silicon absorption features (e.g., \citealt{Maguire2017}). Type Ib \citep{Elias1985} and Ic \citep{Wheeler&Harkness1986} are core collapse SNe (CCSNe), which originate from massive stars ($M > 8 M_\odot$, see \citealt{Smartt2009} for a review), that either do or do not show helium in their spectra, respectively. Stars can lose their hydrogen (and helium) envelopes through, e.g., mass transfer due to binary interactions \citep{Podsiadlowski1992, Eldridge2008} and/or stellar winds \citep{Vink2001, Crowther2007}. Additionally, Type Ic SNe can show very broad lines, called Ic-BL \citep{Modjaz2016, Sahu2018}, or narrow emission lines (Type Icn; \citealt{Fraser2021, Gal-Yam2022, Perley2022}), while Type Ib SNe sometimes show narrow lines (Type Ibn; \citealt{Pastorello2007}).

SNe Type II are also CCSNe, but as mentioned before, their spectra contain hydrogen. The classes of CCSNe are also sub-divided based on their light curves, besides being classified based on spectral properties. Type II SNe are split into Type IIb which have helium rich spectra \citep{Woosley1987}, IIn which show narrow emission lines in their spectra \citep{Schlegel1990}, IIL which show a linear decay in magnitude in their light curves, and IIP which show a plateau of approximately constant brightness in their light curves \citep{Barbon1979, Doggett1985}. In this paper, we focus on Type II CCSNe.

Besides removing (some of) the hydrogen and/or helium envelope, mass loss in massive stars can lead to a dense circumstellar medium (CSM) around a star. When the SN ejecta interact with this CSM, narrow emission lines will appear in the spectrum, creating Type In or IIn SNe as mentioned before. Mass loss in massive stars can arise through different mechanisms. Besides stellar winds and binary interactions, eruptions can release material from the star \citep[e.g.,][]{Smith2014, Yoshida2021}. 
The shape and speed of the CSM combined with the speed and energy of the SN ejecta determines the (late-time) evolution of the light curve and spectra of interacting SNe \citep[e.g.,][]{Dessart2015, Ercolino2024}, see \cite{Dessart2024} for a review of interacting SNe. \cite{Dessart2023} showed that for typical red supergiant (RSG) parameters, CSM interaction would predominantly cause extra brightening in the UV and not as much in the optical and infrared. It is therefore most beneficial to study SNe in the UV to constrain the CSM interaction properties.

The classification of SN type is based on observational properties, which means it is not always clear cut. It can change over time as emission lines appear and disappear due to e.g., interactions with the CSM. For example, \citet{Dong2024} show that the classification of SN2022cvr changes from a IIb to a Ib over time. There are also works that demonstrate that Type IIP and Type IIL SNe are not two distinct types, but rather a continuum of slopes in the plateau due to different amounts of hydrogen being present \citep[e.g.,][]{Anderson2014, Sanders2015, Galbany2016b, Hiramatsu2021}.

With the increase in survey cadence and depth of transient surveys, an increase in the number of peculiar transients has also occurred. For example, luminous fast blue optical transients (\citealp[LFBOTs; e.g,][]{Inserra2019, Metzger2022}) can spectroscopically look similar to SNe, however, their light curves decay on a timescale that is too fast to be powered by the radioactive decay of nickel. Therefore, they are hard to explain with one of the classical SN models. LFBOTs are a class of transients for which there is no clear consensus on the progenitor model. A `naked' engine model in which there is a CCSN with a small ejecta mass, in which the emission is powered by accretion onto a compact object that has formed during the SN, has been proposed \citep[e.g.,][]{Prentice2018, Perley2019, Margutti2019, Mohan2020}. An alternative model is where the tidal disruption of a white dwarf by an intermediate mass black hole is responsible for the emission \citep[e.g.,][]{Kuin2019,Perley2019, Inkenhaag2023}. A third model invokes interaction between an outflow and a dense CSM to explain the observed properties \citep[e.g.,][]{FS2019, Leung2020, Xiang2021, Pellegrino2022a}.  
The prototype of this class of transients is \cow, which was the first of the sample to be discovered in real time \citep{Prentice2018,Perley2019}. It is the most nearby of the sample, and combined with the real time detection, this led to an extensive observing campaign. Despite the resulting rich data set, there is no consensus on the progenitor model of this peculiar transient \citep[see e.g.,][]{Prentice2018, Perley2019, Margutti2019,Xiang2021,Sun2022, Inkenhaag2023}. Other examples of peculiar transients related to SNe are Calcium-strong transients, which are intrinsically faint \citep[e.g.,][]{Kasliwal2012} and super-luminous SNe, which are luminous but rare \citep[e.g.,][]{Quimby2011, Gal-Yam2012}.

The UV light curves of CCSNe decay at a fast rate as the photosphere cools, and the SNe fade away in the UV on a timescale of tens of days \citep{Brown2009, Pritchard2014}. Therefore, if a SN is detected in the UV beyond this timescale, there is a mechanism involved that extends the duration of the UV emission phase, e.g., shock interaction between the SN ejecta and the CSM \citep{Fransson1984, Chevalier1994}. This CSM has been expelled by the progenitor star before the SN explosion, and hence studying the composition of the CSM and the timescale on which the ejecta-CSM interaction happens can help us constrain progenitor models if we detect late-time UV emission. 

The main goal of this work is to investigate the nature of \cow\ through comparing the late time UV emission in CCSNe to that of \cow\ in order to investigate if a CCSN nature of \cow\ is possible. For this, we use a sample of UV  \textit{Hubble Space Telescope (HST)} images obtained within 5 years of the SN discovery, count how many CCSNe are detected in these images and compare the upper limits and detections to the late-time UV emission of \cow. 
Late-time UV and/or optical emission in CCSNe can be an indication of interaction of the SN ejecta with CSM (Type IIn SNe; e.g., \citealp[SN2006gy;][]{Quimby2006, Ofek2007, Smith2007} \citealp[SN2010jl][]{Benetti2010, NP2010, Stoll2011}, \citealp[SN2013L;][]{Monard2013}, \citealp[SN2021adxl;][]{Brennan2023}), so an important secondary goal is to compare our observations to previously known interacting SNe. This paper is observational in nature, we will not focus on modelling the CSM interaction. 

In this paper we investigate what fraction of CCSNe is detected in the UV between 2--5 years after the explosion and compare possible detections to the detected properties of \cow. In Section~\ref{sec:dat_anal} we discuss the sample used and describe the analysis of the sample. This includes obtaining astrometric solutions, measuring the brightness of detected sources and performing an artificial star experiment to determine upper limits in the case of non-detections. In Section~\ref{sec:results} we present the results of our analysis. The results are discussed in the context of interacting SNe in Section~\ref{sec:discussion}. In this section we also discuss what our results mean for the nature of peculiar transient \cow.

In this Manuscript we use H$_0 = 67.8$\,km\,s$^{-1}$\,Mpc$^{-1}$, $\Omega_{\rm m}=0.308$ and $\Omega_{\rm \Lambda} = 0.692$ \citep{Planck2016}. Magnitudes are presented in the AB magnitude system unless specified otherwise and uncertainties are 1$\sigma$ unless stated otherwise.

%--------------------------------------------------------------------
\section{Data analysis}\label{sec:dat_anal}

\subsection{Sample}
\begin{table*}[]
\vspace{-0.2cm}
\caption{Full sample of SNe including relevant information used in this work.}
\label{tab:sample}
\begin{footnotesize}
\begin{tabular}{lcccccccccc}
\hline
Name & RA & Dec & Type & Redshift & $\Delta t^\dagger$& Brightness & A$_{F275W}$ & $\sigma_{WCSsolution}$ & $\sigma_{total}$ & Refs \\
 & (degrees) & (degrees) &  &  & (days) & (mag) & (mag) & (arcsec) & (arcsec) &  \\
\hline 
SN2015bm$^*$&182.00032 & 19.74078 & II & 0.015 & 1777 & >25.2 & 0.14 & 0.2 & 0.25 & (a), (b)\\
SN2016bmi$^*$&278.63403 & -58.52915 & IIP & 0.00746 & 1647 & >25.5 & 0.18 & 0.01 & 0.11 & (c), (d) \\
SN2016bsb&173.61495 & 11.69862 & II & 0.02 & 1699 & >25.5 & 0.45 & 0.037 & 0.05 & (e) \\
SN2016cyx/& \multirow{2}{*}{35.34494} & \multirow{2}{*}{16.56526} & \multirow{2}{*}{II} & \multirow{2}{*}{0.01369} & \multirow{2}{*}{1617} & \multirow{2}{*}{>21.4} & \multirow{2}{*}{0.85} & \multirow{2}{*}{0.025} & \multirow{2}{*}{0.030} & \multirow{2}{*}{(d), (f)} \\
\quad ASASSN-16gy& & & & &  & &  &  &  &  \\

SN2016egz$^*$&1.01621 & -34.81444 & II & 0.0232 & 1762 & >25.55 & 0.19 & 0.0225 & 0.1 & (d), (g) \\
SN2016gkg&23.56016 & -29.44006 & IIb & 0.0049 & 1698& >19.9 & 0.08 & 0.0028 & 0.1 & (d), (h)\\
SN2016hbb&278.94665 & 22.47505 & II & 0.01333 & 1633 & >23.7 & 0.1 & 0.0272 & 0.1 & (d), (i) \\
SN2016hgm/&\multirow{2}{*}{20.54922} & \multirow{2}{*}{0.95206} & \multirow{2}{*}{II} & \multirow{2}{*}{0.008} & \multirow{2}{*}{1447} & \multirow{2}{*}{>25.4} & \multirow{2}{*}{0.29} & \multirow{2}{*}{0.1225} & \multirow{2}{*}{0.13} & \multirow{2}{*}{(j)} \\
\quad SNHUNT327& & & & & & & & & & \\
SN2016hvu&338.9815 & 20.32016 & IIP & 0.01852 & 1452 & >25.5 & 0.89 & 0.042 & 0.11 & (d), (k) \\
SN2016i/&\multirow{2}{*}{219.93654} & \multirow{2}{*}{23.39514} & \multirow{2}{*}{IIP} & \multirow{2}{*}{0.01490} & \multirow{2}{*}{1596} & \multirow{2}{*}{>23.7} & \multirow{2}{*}{0.17} & \multirow{2}{*}{0.1} & \multirow{2}{*}{1} & \multirow{2}{*}{(d), (l)} \\
\quad ASASSN-16ai$^{**}$& & & & & & & & & &  \\
SN2016jbu&114.10815 & -69.54868 & IIn & 0.00489 & 1553 & >23.45 & 0.25 & 0.0196 & 0.1 & (m), (n) \\
SN2017bif&261.613 & -60.54437 & II & 0.019247 & 1342 & >25.3 & 1.1 & 0.0242 & 0.1 & (o), (p)\\
SN2017caw&121.95875 & -61.771 & II & 0.027145 & 1505 & >24.45 & 0.52 & 0.043 & 1 & (q)\\
SN2017dch&298.89873 & -11.63964 & Ic & 0.048 & 1499 & >25.45 & 1.04 & 0.0168 & 0.1 & (r)\\
SN2017dek&306.4363 & -21.66108 & II & 0.063 & 1627 & >25.5 & 0.88 & 0.0241 & 0.1 & (s)\\
SN2017dhu&79.13307 & 6.46287 & II & 0.043 & 1620 & >25.4 & 0.34 & 0.2 & 0.22 & (t)\\
SN2017dka&210.44193 & 9.49926 & II & 0.021 & 1820 & >25.0 & 0.43 & 0.0521 & 0.11 & (u)\\
SN2017dkb&273.51429 & 21.87742 & IIP & 0.016 & 1604 & >25.4 & 1.0 & 0.046 & 1 & (v)\\
SN2017eca&241.06547 & 17.95765 & IIP & 0.032 & 1598 & >25.5 & 0.14 & 0.044 & 0.11 & (w) \\
SN2017eiy&357.36779 & -30.41797 & IIb & 0.047 & 1547 & >25.45 & 0.62 & 0.2 & 0.22 & (w)\\
SN2017faa/&\multirow{2}{*}{199.76626} & \multirow{2}{*}{-2.51275} & \multirow{2}{*}{II} & \multirow{2}{*}{0.018480} & \multirow{2}{*}{1434} & \multirow{2}{*}{>23.8} & \multirow{2}{*}{0.29} & \multirow{2}{*}{0.053} & \multirow{2}{*}{0.12} & \multirow{2}{*}{(p), (x)} \\
\quad ATLAS17hpc& & & & & & & & &  \\
SN2017fbu&32.77892 & 3.84349 & II & 0.01086 & 1195 & >25.45 & 0.27 & 0.168 & 0.17 & (p), (y)\\
SN2017ffm&349.59592 & -4.41619 & II & 0.024 & 1521 & >25.4 & 0.09 & 0.084 & 0.09 & (z)\\
SN2017ffq&265.06057 & -58.43244 & II & 0.01272 & 1302 & >25.4 & 0.22 & 0.0238 & 0.1 & (p), (aa)\\
SN2017fgk&266.95498 & 16.13477 & Ic-BL & 0.034 & 1317 & >25.45 & 0.25 & 0.01 & 0.023 & (ab)\\
SN2017fod&281.23116 & 27.30721 & II & 0.045 & 1521 & >22.55 & 0.54 & 0.025 & 0.03 & (ac)\\
SN2017fqk$^{**}$&43.50871 & 2.96881 & II & 0.01015 & 1195 & >22.85 & 0.52 & 0.156 & 1 & (p), (ad) \\
SN2017fqo&28.25629 & 12.71279 & II & 0.015 & 1530 & >24.35 & 0.82 & 0.2 & 0.22 & (ae) \\
SN2017fvr&71.86761 & 23.98273 & IIP & 0.013 & 1222 & >25.45 & 0.51 & 0.0162 & 0.1 & (af)\\
SN2017fwm&288.21651 & -60.38288 & Ic & 0.016 & 1396 & >25.45 & 0.36 & 0.0202 & 0.1 & (ag)\\
SN2017ggw&37.61203 & -43.01471 & II & 0.018 & 1353 & >25.4 & 4.39 & 0.051 & 0.11 & (ah)\\
SN2017gip$^*$&346.25255 & 28.76174 & Ic & 0.031 & 1145 & >23.5 & 0.33 & 0.14 & 0.17 & (ai)\\
SN2017giq&359.47811 & 28.50335 & Ic & 0.029813 & 1139 & >25.45 & 0.12 & 0.0257 & 0.03 & (aj)\\
SN2017gry/&\multirow{2}{*}{52.03309} & \multirow{2}{*}{-56.57833} & \multirow{2}{*}{IIP} & \multirow{2}{*}{0.019337} & \multirow{2}{*}{1337} & \multirow{2}{*}{22.04$\pm$0.01} & \multirow{2}{*}{0.19} & \multirow{2}{*}{0.2} & \multirow{2}{*}{0.22} & \multirow{2}{*}{(p), (ak)}\\
\quad GAIA17chn& & & & & & & & &  \\
SN2017hcc/&\multirow{2}{*}{0.96079} & \multirow{2}{*}{-11.47461} & \multirow{2}{*}{IIn} & \multirow{2}{*}{0.0173} & \multirow{2}{*}{1421} & \multirow{2}{*}{>24.65} & \multirow{2}{*}{0.16} & \multirow{2}{*}{0.0825624} & \multirow{2}{*}{0.13} & \multirow{2}{*}{(p), (al)}\\
\quad ATLAS17lsn& & & & & & & & & &  \\
SN2017hxv/&\multirow{2}{*}{326.09567} & \multirow{2}{*}{-29.91649} & \multirow{2}{*}{II} & \multirow{2}{*}{0.01874} & \multirow{2}{*}{1130} & \multirow{2}{*}{>25.45} & \multirow{2}{*}{0.23} & \multirow{2}{*}{0.0192} & \multirow{2}{*}{0.13} & \multirow{2}{*}{(p), (am)}\\
\quad ASASSN-17oj$^*$& & &  & & & & & &  \\
SN2017ipa&99.49229 & -15.36683 & II & 0.029 & 1239 & >25.5 & 0.46 & 0.0113 & 0.1 & (an)\\
SN2017ivu&234.13629 & 16.60558 & IIP & 0.006528 & 1267 & >25.35 & 0.34 & 0.0151 & 0.1 & (p), (ao)\\
SN2017ivv/&\multirow{2}{*}{307.20742} & \multirow{2}{*}{-4.38261} & \multirow{2}{*}{II} & \multirow{2}{*}{0.0056} & \multirow{2}{*}{1252} & \multirow{2}{*}{25.57$\pm$0.06} & \multirow{2}{*}{0.31} & \multirow{2}{*}{0.018} & \multirow{2}{*}{0.028} & \multirow{2}{*}{(ap), (aq)} \\
\quad ASASSN-17qp& & & & & & & & & &  \\
SN2017jbj$^*$&12.02255 & -2.78958 & II & 0.013492 & 1384 & >25.5 & 2.09 & 0.0489 & 0.028 & (p), (ar)\\
SN2017mw&149.33736 & -41.58916 & IIb & 0.012 & 1655 & >25.35 & 0.26 & 0.025 & 0.1 & (as)\\
SN2018aad/&\multirow{2}{*}{59.50643} & \multirow{2}{*}{-65.50676} & \multirow{2}{*}{II-pec} & \multirow{2}{*}{0.025} & \multirow{2}{*}{1155} & \multirow{2}{*}{>25.5} & \multirow{2}{*}{1.09} & \multirow{2}{*}{0.2} & \multirow{2}{*}{0.22} & \multirow{2}{*}{(at), (au)}\\
\quad ASASSN-18eo& & & & & & & & &  \\
SN2018ant&129.13103 & -11.82802 & II & 0.0197 & 952 & >25.2 & 0.23 & 0.01 & 0.1 & (au), (av)\\
SN2018anu&264.05979 & 18.98239 & II & 0.039 & 1538 & >25.45 & 0.42 & 0.0436 & 0.08 & (aw)\\
SN2018dfg&211.64462 & -5.45049 & IIb & 0.00948 & 953 & >25.45 & 1.42 & 0.0341 & 0.08 &(au), (ax) \\
SN2018eog&307.05021 & -3.13633 & II & 0.02 & 970 & >24.6 & 0.28 & 0.016 & 0.1 & (au), (ay)\\
SN2018evy&275.65907 & 15.69657 & II & 0.01757 & 959 & >25.55 & 0.35 & 0.0258 & 0.1 & (au), (az)\\
SN2018fit&351.31739 & 13.93406 & II & 0.014 & 1001 & >24.6 & 0.15 & 0.0205 & 0.08 & (ba), (ba)\\
SN2018iuq&106.47264 & 12.89296 & IIb & 0.028 & 887 & >25.2 & 0.41 & 0.1038 & 0.14 & (au), (bb)\\
SN2018jfz&50.07807 & -0.29633 & II & 0.034 & 732 & >25.0 & 1.17 & 0.2 & 0.2 & (bc) \\
SN2018pq/&\multirow{2}{*}{193.88019} & \multirow{2}{*}{-50.05472} & \multirow{2}{*}{II} & \multirow{2}{*}{0.00711} & \multirow{2}{*}{1654} & \multirow{2}{*}{>25.05} & \multirow{2}{*}{0.17} & \multirow{2}{*}{0.027} & \multirow{2}{*}{0.12} & \multirow{2}{*}{(au), (bd)}\\
\quad ASASSN-18cb$^*$& & & & & & & & & &  \\
\hline 
\end{tabular}
\newline $^\dagger$ Time between discovery of the SN and the observations used in this work.
\newline $^*$ For these SNe we determined the position of the SN as described in Section~\ref{sec:manual_alignment} .
\newline $^{**}$ These SNe were reported by ASAS-SN and we found no image with a SN detection to manually determine the SN position and accuracy. See section~\ref{sec:manual_alignment} for details on how the position was obtained.
\newline $^\ddagger$ See after acknowledgements for list of references.
\end{footnotesize}
\end{table*}

Our sample consists of 51 nearby (z<0.065) CCSNe observed with the \textit{HST} within five years of their discovery (see Table~\ref{tab:sample}). This sample is a subset of a public UV snapshot survey of CCSNe environments (Proposal 16287; PI J.D.~Lyman), to compliment observations done in the All-weather MUse Supernova Integral field Nearby Galaxies (AMUSING) survey. This survey is a long-term programme using the Very Large Telescope (VLT)/Multi Unit Spectroscopic Explorer (MUSE) to obtain integral field observations of SN host galaxies in suboptimal atmospheric conditions \citep{Galbany2016a}. The AMUSING sample serves as a statistical
% ly representative 
sample to study nearby SN host galaxies. We select the sources from the \textit{HST} snapshot survey that have been observed within five years of discovery, which allows us to compare the results on the timescale of the observations of \cow\ used by \cite{Sun2022} and \cite{Inkenhaag2023}. 
We use the redshift listed in Table~\ref{tab:sample} and the cosmological parameters mentioned in the introduction to calculate the luminosity distance using the {\sc Distance} package within {\sc astropy.cosmology}.  
For sources with z<0.01 (eight in total) we check if the distance listed in the NASA/IPAC Extragalactic Database (NED)\footnote{\url{https://ned.ipac.caltech.edu/}} or the literature is the same as our calculated distance. For these eight sources there is either no distance listed in NED or the literature, or the value is consistent with our calculated values within 3$\sigma$. We therefore proceed with our calculated distances for all sources.

The observations were done in one filter (F275W) per source with a three or four point drizzle. Depending on the position on the sky, this results in an exposure time between 1350 and 1800 seconds per source. 
We downloaded the reduced final products (\texttt{\_drc} images) from the \textit{HST} archive, which are the combined, drizzled images that have been calibrated and corrected for geometric distortion and charge transfer efficiency.

\subsection{Astrometry}

The first step in our analysis is to improve the astrometry of the images to obtain the positions of the SNe as accurately as possible, including determining the full astrometric uncertainty in the position of the SNe. This is to make sure that we are not misidentifying detected UV point source emission, for instance from a nearby star forming (SF) region, as coming from the SN.

\subsubsection{Alignment of images}\label{sec:align}

Whilst the world coordinate system (WCS) solution in the header of {\it HST} images is generally good with an error <0.2 arcsec \footnote{\url{https://outerspace.stsci.edu/display/HAdP/Improvements+in+HST+Astrometry}\label{note1}}, we re-align the images using the WCS position of {\it Gaia} sources to increase the WCS accuracy. {\it HST} images that are already aligned to {\it Gaia} through the Guide Star Catalog have not be re-aligned, as their WCS solution is already tied to \textit{Gaia}. For these images, the error in the WCS solution is taken to be 0.01~arcsec \footnote{See footnote \ref{note1}}.

To correct the astrometric solution we started by using {\sc astroquery} \citep{astroquery} to download all the sources in \textit{Gaia} data release 2 (DR2) within 2\arcmin of the source location. We then compared the \textit{HST} image with the list of sources from \textit{Gaia} DR2 and delete any of the sources that are not visible by eye in the {\it HST} image. Then we used {\sc ccmap} in {\sc iraf} \citep{iraf1, iraf2} to make a new WCS solution based on the \textit{Gaia} coordinates and the pixel coordinates in the \textit{HST} image (obtained using {\sc imexam} in {\sc iraf}). During this procedure with {\sc ccmap} we use \texttt{fitgeom=`general'} for images with more than 5 sources, while for images with 3 to 5 sources we use \texttt{fitgeom=`rscale'} in order to avoid over-fitting the available tie points. For images where there were only one or two Gaia sources detected, we consult the PanSTARRS1 (PS1) DR2 database to check if there are more PS1 sources detected in the \textit{HST} image and if so, we use the PS1 DR2 database to compute a new WCS solution.

Images with only one or two sources in either the PS1 DR2 or Gaia DR2 database were aligned using the {\sc Gaia} software \citep{Gaia_software}. The catalogue with the most matching sources ({\it Gaia} or PS1) was used for this alignment, preferring Gaia over PS1 when the number of sources is the same in both because the \textit{Gaia} astrometric uncertainty is smaller. We used the {\sc Astrometric Calibration} function under {\sc Image-Analysis}. Since the rotation of {\it HST} is well known, it is possible to align images using one or two sources by only adding shifts in the x and y direction. This was done using {\sc Tweak an existing calibration}. In the cases where two sources were detected, the RMS on the fit was used as the error in the astrometry. If there was only one source to align the WCS solution we use the largest error of 0.2\arcsec mentioned in the 06-2022 "Instrument Science Report WFC3" \footnote{See footnote \ref{note1}}. In practice, this is a highly conservative estimate since the dominant source of error in this case is the uncertainty of the source positions in the Gaia (or PS) frames, which are likely $< 0.1\arcsec$. After the WCS solution was updated in the header (either using the {\sc iraf} or the {\sc Gaia} software), we performed a visual confirmation to ensure the procedure was executed correctly. The new WCS solution changes the SN position by up to $\sim$10~pixels (or $\sim$0.4~arcsec) per image.

\subsubsection{Manual determination of SN positions} \label{sec:manual_alignment}

Four SNe used in this work were reported by amateur astronomers and not by any other survey/telescope, which means they are not reported including an error on the position (SN2015bm, SN2017bmi, SN2017caw and SN2017jbj). Four sources are only reported by ASAS-SN (SN2016egz, SN2017fqk, ASASSN-17oj and ASASSN-18cb), which has an uncertainty of 1.0\arcsec\ (see Table~\ref{tab:accuracies}). Two additional sources (SN2017dkb and SN2017gip) are only reported by the PMO-Tsinghua Supernova Survey (PTSS), for which no accuracy is reported either. We first check if there are any point sources detected within 3$\sigma$ if we set the $1\sigma$ positional accuracy of the SN position to 1\arcsec. Any error we will obtain on the position of the SN is likely smaller than this 1\arcsec, so there is no need to improve the accuracy if no point sources are detected within $3\sigma$ using a 1\arcsec error region. This is the case for SN2017dkb, for which we thus do not follow the procedure described below. 
For the remaining nine sources, we aim to obtain or improve the accuracy on the position of the SN. 
To obtain the accuracy of the position we searched observatory archives for observations where the SN was detected. For three SNe (ASASSN-16ai, SN2017caw and SN2017fqk) we do not find any archival images where the SNe were detected. For these sources we use a positional error of 1\arcsec and for ASASSN-16ai and SN2017fqk we use the reported position from \cite{Holoien2017b} and \cite{Holoien2019}, respectively, which are confirmed to come from follow-up observations so using a 1$\sigma$ uncertainty is warranted. Five SNe (ASASSN-17oj, SN2015bm, SN2016bmi, SN2016egz, SN2017gip and ASASSN-18cb) have acquisition images from their classification spectra in the European Southern Observatory (ESO) archive. For the remaining source, SN2017jbj, the classification was done by the Padova-Asiago group, from whose archive\footnote{\url{http://archives.ia2.inaf.it/aao/}} we downloaded the sloan g-band images taken right before the classification spectrum was taken. None of the (acquisition) images contain a WCS solution in the header, so to obtain an initial solution we use {\sc astrometry.net}\footnote{\url{astrometry.net}}. 
Next, we followed the same procedure to obtain a better WCS solution for these images with a SN detection as mentioned above in Section~\ref{sec:align} for the \textit{HST} images. The images with the SN detections have a larger field of view than the \textit{HST} images, so we used all \textit{Gaia} sources visible by eye in the images within 10\arcmin\ instead of 2\arcmin. We then measured the position of the SN on this aligned image and used the RMS on the new WCS solution of the images with a detection of the SNe as a measure of the uncertainty of the position. 

\subsubsection{Proper motion}

We check the importance of including the proper motion of the \textit{Gaia} stars between the observational epoch of \textit{Gaia} DR2 (J2015.5) and the time of observations of our \textit{HST} data. We perform this check for the images we aligned to the Gaia DR2 database, but we are not able to perform this check for the images aligned to the PS1 DR2 database as no proper motion measurements are included in this data release.
Using {\sc astroquery}, we obtain the values of the proper motions in RA and Dec for the sources we used to perform the astrometry. We remove sources with a value less than three times the error on the value. This way we only use sources with a proper motion inconsistent with zero within $3\sigma$. We then add all the values for RA and all values for Dec separately and divide these total values by the amount of sources left after only using 3$\sigma$ significant proper motion values. For all but two images this calculation results in an average proper motion effect of less than two-and-a-half pixels (0.1~arcsecs) in RA and Dec direction. For SN2017gip and GAIA17chn there is an average proper motion in either RA or Dec direction that is bigger than two-and-a-half pixels (0.1~arcsecs), but still smaller than five-and-a-half pixels (0.22~arcsecs). There are no sources around any of the SN positions that are outside the 3$\sigma$ uncertainty regions but would be inside this uncertainty region if the effect of proper motion was accounted for. Proper motion effects therefore do not influence our conclusions.

\subsubsection{Full astrometric uncertainty}

The total error on the astrometry ($\sigma_{total}$) is calculated as follow:
\begin{equation}
    \sigma_{total} = \sqrt{\sigma_{survey}^2+\sigma_{WCS solution}^2+\sigma_{SN position}^2}\label{eq:unc}
\end{equation}
where $\sigma_{survey}$ is the positional accuracy of the survey used to align the {\it HST} images (either 0.021~arcsec if PS1; \citealt{PS_acc}, or 0.002~arcsec if {\it Gaia}\footnote{\url{https://www.cosmos.esa.int/web/gaia/dr2}}), $\sigma_{WCS solution}$ is the error on the new WCS solution of the \textit{HST} image determined in the way as described in Section~\ref{sec:align}, and $\sigma_{SN position}$ is the error on the reported position of the SN. Table~\ref{tab:accuracies} contains the astrometric uncertainty of the surveys/telescopes that reported SNe within the sample (which represents the error on the reported position of the SNe if not determined manually as described above in Section~\ref{sec:manual_alignment}). If multiple surveys have reported the position of a SN, we use the position reported by the survey with the smallest astrometric uncertainty. {Throughout this work, when we mention the uncertainty on the position of a SN, we mean the full uncertainty as calculated by Equation~\ref{eq:unc} and listed in the last column of Table~\ref{tab:sample}, unless specified otherwise.}

We do not include the centroiding error on the position of the SNe in the detection images, which is defined as $\sigma_{centroid} = FWHM/(SNR*2.35)$ in arcseconds (for the Full-Width at Half Maximum (FWHM) in arcseconds). Since not all surveys release the SN discovery images, it is not possible for us to measure the FWHM and the signal-to-noise ratio (SNR) to {determine} the centroiding error. Assuming FWHM $\approx1$\arcsec and a SNR>10, $\sigma_{centroid}<<0.1$\arcsec, which is smaller than the dominating factor in the error calculation. We therefore do not include this error in the total error budget calculation for the astrometry, and we emphasise that therefore the total error on the position we report is most likely an underestimation of the real error on the position of the SNe. However, the SNe in the sample are typically (bright) nearby SNe, so this unaccounted for error is not likely to dominate.

\begin{table}
\centering 
\caption{Errors on the reported positions of supernovae for various surveys and telescopes that have reported SNe used in this paper.} 
\label{tab:accuracies}
\hspace*{-.2cm}\begin{tabular}{lcc}
\hline
Survey & $\sigma_{SN position}$ & reference \\
 & (arcsec) & \\
\hline %
ASASSN & <1.0 & \cite{ASASSN_acc} \\
Gaia alerts & 0.1 & \cite{Gaia_alerts_acc} \\
CRTS & 0.1$^\dagger$ & \cite{CRTS_acc} \\
Pan-STARRS1 & 0.021$^*$ & \cite{PS_acc} \\
ATLAS & 0.07 & \cite{ATLAS_acc} \\
\hline
\end{tabular}
\newline $^\dagger$ CRTS does not give an accuracy, but they report declinations to 1 decimal, so we use 0.1 arcsec as the uncertainty \\
$^*$ Pan-STARRS1 reports an RMS of 0.016~arcsec and a systematic uncertainty of 0.005~arcsec with respect to {\it Gaia}, so added linearly this becomes 0.021~arcsec. \\
\end{table}

\subsection{Source detection} \label{sec:Pch}

We ran {\sc Source Extractor} \citep{Bertin1996} on all WCS-corrected \textit{HST} images to obtain a list of sources in the images. {{\sc Source Extractor} is computationally cheap, but still accurate for source detection on non-crowded images such as our UV \textit{HST} images.} We used default parameters and \texttt{DETECT\_THRESH = 3.0} and \texttt{DETECT\_MINAREA = 5}. These parameter settings minimise the chance of detecting artefacts such as hot pixels. 

We checked if any detected sources lie within 1, 2 or 3$\sigma$ uncertainty of the position of the supernova.
Next, we split the sample into three groups: \textit{i)} images where there is no underlying emission (point sources or extended sources) within the uncertainty of the SN position, \textit{ii)} images where there is extended emission overlapping with (or close to) the positional uncertainty of the SN and \textit{iii)} images where this is a point source within the positional uncertainty of the SN. For this last category we only report sources that are detected when performing point spread function (PSF) photometry with {\sc DOLPHOT} as described in Section~\ref{sec:phot} below and setting \texttt{threshold=5}. The latter threshold value is set to 5 to minimise the chance of reporting background Poisson fluctuations as potential detections. 

\subsection{Photometry}\label{sec:phot}

We perform PSF-photometry on the images where we found a point source at the SN location to obtain its magnitude. 
{ We switch from {\sc source extractor} which can only perform aperture photometry, to }{\sc DOLPHOT} (v2.0; \citealt{Dolphin2000}), which is specifically designed to perform PSF photometry on \textit{HST} data. The photometry is performed on the individual \texttt{\_flc} images and then combined into one (Vega) magnitude per source. We convert these Vega magnitudes to AB magnitudes using the difference in zero-points from the WFC3 handbook\footnote{\url{https://www.stsci.edu/hst/instrumentation/wfc3/data-analysis/photometric-calibration/uvis-photometric-calibration}}. The parameter files for {\sc DOLPHOT} are provided in the reproduction package of this manuscript \url{Zenodo url}.

\subsection{Chance alignment probability}

For images where there is a point source within the uncertainty region of the SN we calculate the chance alignment of emission {within the circular region with the distance between the galaxy centre and the SN position as radius}. {This is equivalent to calculating the probability of a random source caused by unrelated UV emission in the host galaxy coinciding with the SN position.} 
The host of ASASSN-17qp/SN2017ivv is GALEXASC~J202849.46-042255.5 with RA: 307.206100, Dec: -4.382107~deg and the host of ATLAS17lsn/SN2017hcc is WISEA~J000350.27-112828.7 with RA: 0.959520, Dec:-11.474532~deg\footnote{The coordinates were obtained from the NED database: \url{https://ned.ipac.caltech.edu/}}.

We map this region to the object pixel mask of the \textit{HST} image produced by {\sc source extractor}.
The object pixel mask is an image in which all pixels that are not considered part of a detected object are 0, while the pixels that form objects retain their value. We count the pixels in the circular region that are not 0 and divide this by the total amount of pixels in the circular region to obtain the chance of the SN position falling on an illuminated pixel in the galaxy. 

\subsection{Magnitude upper limits}

To determine upper limits on emission from the CCSNe in the sample we determine the completeness limit and the limiting magnitude for all {\it HST} images by performing an artificial star experiment. In this experiment we add an artificial star close to the position of the SN on the final drizzled \texttt{\_drc} \textit{HST} image with the new WCS solution and investigate if this artificial star is recovered by our detection algorithm. For details of the procedure see \cite{Eappachen2022}. As in \cite{Eappachen2022} we define the completeness limit as the magnitude at which 95 percent of artificial stars are recovered at >5$\sigma$ within 0.3 magnitude of the assigned magnitude and the limiting magnitude where 33 percent of the artificial stars are recovered.  We use a region with a radius of $3\sigma$ positional accuracy for each SN location in the experiment and magnitude bins of 0.05 magnitude. This experiment also naturally accounts for only brighter point sources being recovered when extended emission is present within the $3\sigma$ uncertainty region of a SN.

For this artificial star experiment, we construct a PSF model for the {\it HST} image of SN2017ffq using {\sc iraf}. This image has sufficient sources spread out across the field of view to construct a representative PSF model. To check if this PSF model is suitable for use on all {\it HST} images, we perform PSF photometry on a randomly selected {\it HST} image from our SN sample and subtract the PSF for all detected sources. The subtraction leaves no residual at the position of the subtracted PSFs, therefore we deem it appropriate to use one PSF model for all images. We use {\sc iraf} for this artificial star experiment instead of {\sc DOLPHOT} because the latter is more computationally expensive and for the purpose of determining the upper limit the computed {\sc iraf} PSF is accurate enough.

We use the completeness limit of each image together with the header keywords regarding the zero point (\texttt{PHOTZPT}), inverse sensitivity (\texttt{PHOTFLAM}) and pivoting wavelength (\texttt{PHOTPLAM}) as described in the WFC3 data handbook\footnote{\url{https://hst-docs.stsci.edu/wfc3dhb/chapter-9-wfc3-data-analysis/9-1-photometry}} to determine the 95 percent confidence upper limit on the magnitude of any emission at the position of the SNe. We assume no host galaxy extinction and correct for Galactic extinction using the {\sc python} package {\sc gdpyc}, with the \cite{SFD1998} dust map to obtain E(B-V) colours and the E(B - V) to extinction conversion coefficients estimated in \cite{S&F2011} for this dustmap to calculate the extinction correction in the {\it HST} WFC3/F275W filter.

\section{Results} \label{sec:results}

\begin{figure*}
 \centering
    \includegraphics[width=.8\textwidth]{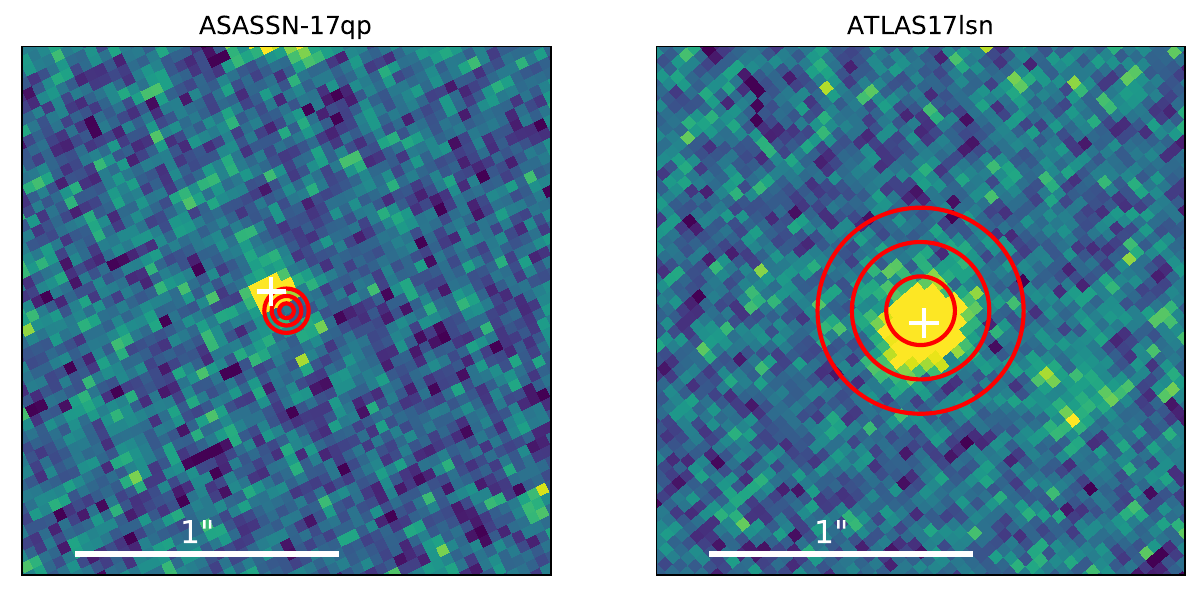}
    \caption{$2\arcsec \times 2\arcsec$ cutout images centred on the positions of the two SNe for which we detect a point source close to the SN position. ASASSN-17qp is displayed on the left and ATLAS17lsn on the right. In red the 1, 2 and 3$\sigma$ uncertainties on the position of the SN are indicated. The centroid positions of the point sources that we considered to be potential SN detections are marked by white crosses. For ASSASN-17qp the white cross is just outside the 3$\sigma$ uncertainty region, however, in Section~\ref{sec:pos_detections} we discuss why we still consider this a possible detection of the SN. In both images North is up and East is left.}
 \label{fig:poss_det}
\end{figure*} 

In our sample of 51 SNe we find two cases where there is a potential detection of the SN (at 5$\sigma$ confidence level) within the 3$\sigma$ uncertainty region of the SN position. Figure~\ref{fig:poss_det} shows the 1, 2, and 3$\sigma$ positional error on the SN location of the two SNe where there is a point source detected within the 3$\sigma$ uncertainty region. For ASASSN-17qp the potential SN detection is just outside the 3$\sigma$ uncertainty region at 3.2$\sigma$; we will discuss below (in Section~\ref{sec:17qp}) why we still consider this point source related to the SN. 

For 36 SNe no point source is detected within 3$\sigma$ uncertainty of the SN position and for 13 SNe we detect extended emission within (or close to) $3\sigma$ of the SN position but there is no detection of a point source on top of this extended emission. It is plausible that faint sources could be confused within the extended emission. However, our artificial star experiment accounts for this, and our limits for supernovae which lie on extended emission are correspondingly brighter. Table~\ref{tab:sample} lists the full sample of SNe including upper limits for sources with non-detections and Figure~\ref{fig:SN_mosaic} shows a 1\arcsec or 4\arcsec cutout centred on the SN position with the 1, 2 and 3$\sigma$ uncertainty region indicated in red circles.

The two SNe for which we possibly still detect emission related to the SN are ASASSN-17qp/SN2017ivv and ATLAS17lsn/SN2017hcc. 
Both of these supernovae have previously been identified as interacting events \citep{Gutierrez2020, Smith&Andrews2020}, and so they are amongst the most likely systems for late-time UV detections. In Figure~\ref{fig:SN_mosaic} there is also a faint detection within the error region of SN2018ant. We looked at the individual \texttt{\_flc} images to check if the source is an artefact from cosmic ray removal before combining the images. In the individual images there is a cosmic ray at/near the position of the detection in two of the three images and in the third one there is no source visible (see Figure~\ref{appfig:18ant_flc}). We also use {\sc astrodrizzle} to re-drizzle and combine the images, using \texttt{driz\_cr\_grow = 3} for better cosmic ray removal. This produces an image without detection of a point source in the uncertainty region of the SN position. There has also been no mention of this SN as a long-term emitter in the literature. For SN2018ant we therefore assume this faint detection is not a real source.

\begin{figure*}
 \centering
    \includegraphics[width=.75\textwidth]{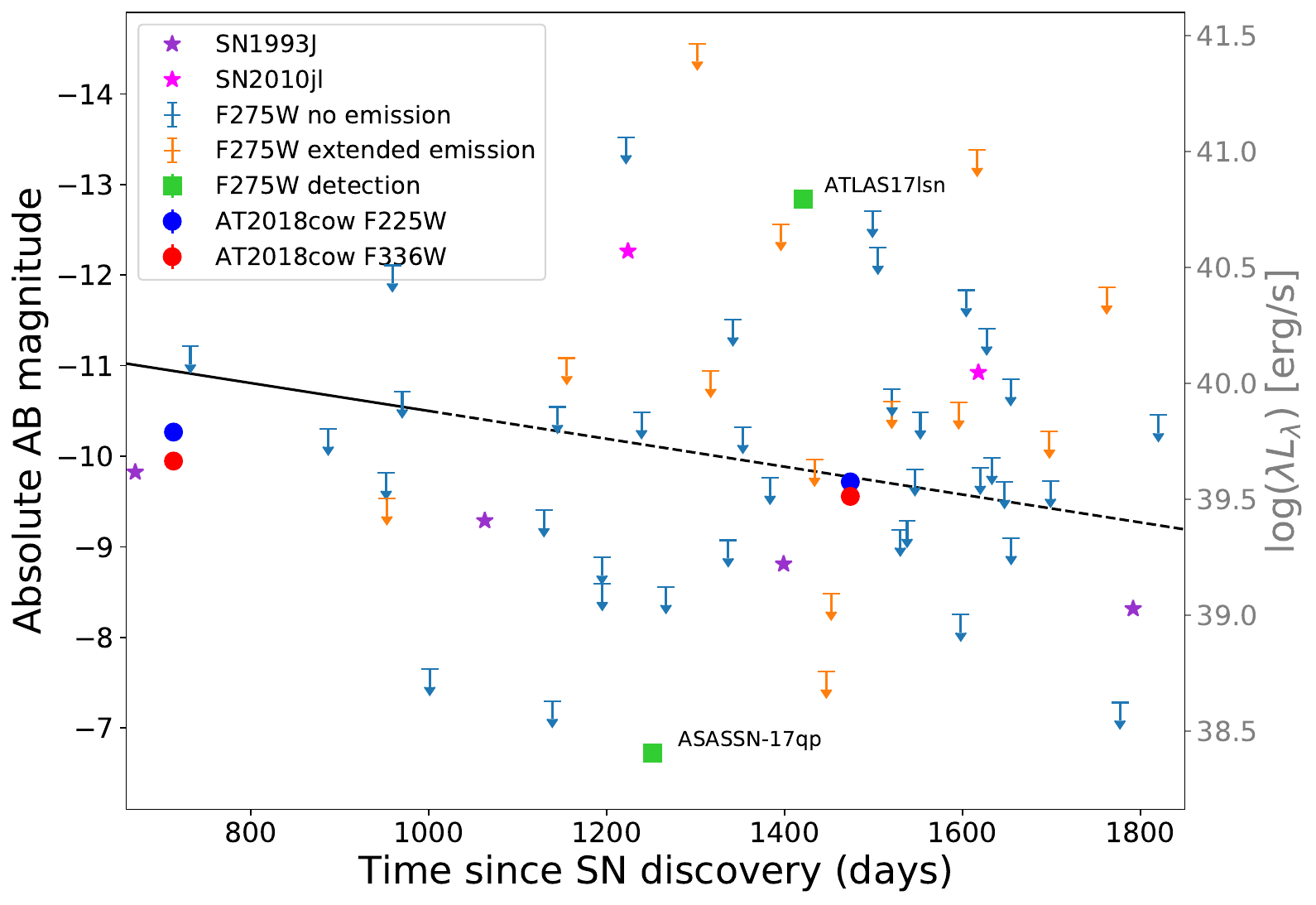}
    \caption{The brightness of our sample of SNe versus time since discovery. The light curve of AT2018cow from \cite{Inkenhaag2023} at late times
    is plotted as well for comparison with circular markers in blue (F225W) 
    and red (F336W). The black line represent the late time UV (\textit{UVW2}-filter) brightness of the model with CSM interaction from \cite{Dessart2023}, and the dashed line is the model extrapolated with the same decay rate for the duration of our observations. 
    The light blue arrows represent upper limits to SNe emission determined close to/at the position of the SN, while orange arrows represent upper limits to SNe where there is extended emission close to/at the position of the SN. Light green squares indicate the magnitude of the detection of a point source in the UV at late times at a position consistent with that of the SN. They are labelled with the corresponding SN name and the error on the magnitude is smaller than the marker size. The magenta and purple stars are synthetic photometric measurements of SN2010jl and SN1993J, respectively, see the main text for details on how these were obtained. On the right axis we have added the shock luminosity of the SN in erg~s$^{-1}$ corresponding to the absolute magnitude on the left axis for clarity. 
    }
 \label{fig:UL_lightcurve}
\end{figure*}

Figure~\ref{fig:UL_lightcurve} shows the upper limits on the detection of a UV counterpart to the sample of SNe as a function of time since discovery. The light blue markers represent SNe where we detect no extended emission around/at the position of the SN, while the orange markers represent SNe where extended emission around the position of the SNe was detected in the {\it HST} images. While the events with extended diffuse emission typically yield brighter limiting magnitudes, the range of distances is such that this is not obvious in the absolute magnitude distribution. SNe for which a point source was detected inside the 3$\sigma$ positional uncertainty region of the SN are marked by green squares with error bars. For comparison, the light curve of \cow\ is also plotted in two filters close to the F275W filter used on this work.

\section{Discussion} \label{sec:discussion}

We do not detect a point source within the 3$\sigma$ uncertainty region for 49 of our 51 SNe, which corresponds to 96~percent of the cases. For 2 out of 51 (4~percent) we do find a point source that could be related to late-time emission from the CCSN. 

For 13 of the 49 SNe where we do not find a point source, there is evidence for the presence of extended emission close to the SN position. As expected, the artificial star experiment demonstrates that point sources need to be brighter to be recovered, leading to a less deep magnitude upper limit when extended emission is present. Additionally, for the two cases with a point source detected we do not find extended emission close to the SN position. At the distances of ASASSN-17qp and ATLAS17lsn the PSF size of 1.9~pixels in F275W corresponds to size of $\sim9$ and $\sim29$~parsec, respectively. 
Combined, this means that a SF region is only detected at, or at close projection of, the position of 25 percent of the SNe in our sample (assuming all UV emission is from SF regions). This assumption is reasonable because emission in the UV is dominated by young, massive stars which are born in SF regions. However, the luminosity function of SF regions is steep ($\alpha \lesssim -1.73$ \citealt[e.g.][]{Cook2016, Santoro2022}) and therefore we cannot detect all the SF regions in the SN host galaxies due to the distance to the host galaxy. Hence, we can say our calculated 25~percent of CCSNe near a SF region is a lower limit to this number, and we refrain from making comparisons to other works investigating SN associations with SF regions which are typically made on much more local samples.  

To get a better understanding of the progenitors of CCSNe and the nature of the peculiar transients recently discovered, one can also study their environments \citep[e.g.,][]{Kuncarayakti2013a, Kuncarayakti2013b, Galbany2016a, Anderson2015, Kuncarayakti2018, Lyman2020}. Studying the environments in UV specifically, provides a window into star formation and massive star populations. As we do not detect a point source in most of our images on timescales between two and five years after the discovery of the SNe, we conclude that this is suitable timescale to consider if investigating the local environment is the goal.

\subsection{Point source detections}\label{sec:pos_detections}

The two SNe for which we detect a point source are ASASSN-17qp/SN2017ivv (Type II) and ATLAS17lsn/SN2017hcc (Type IIn). Both of these SNe are hydrogen-rich and likely had a massive progenitor that has shed part of its envelope through stellar winds, creating CSM with which the SN ejecta can interact. This interaction is primarily visible through (enhanced) UV emission \citep{Dessart2022}. We will discuss both cases individually below.

\subsubsection{SN2017hcc/ATLAS17lsn}
From Figure~\ref{fig:poss_det} one can see that ATLAS17lsn is detected in the 1$\sigma$ uncertainty region. The source has a chance of a false positive due to the SN position coinciding with pre-existing emission out to the offset of the SN of ${1.5\times10^{-3}}$.

\cite{Moran2023} report emission for this source from the optical through to the infrared (IR) for more than 1700 days after its discovery. They argue this late time emission is due to interaction between massive, dense CSM and the SN ejecta, creating a slow rise to peak and a very slow, long decay of the light curve. This SN is known to most likely have had a massive progenitor that lost $\sim10$~M$_\odot$ of mass before the explosion, creating shells of circumstellar material with which the supernova ejecta is interacting (\citealt{Smith&Andrews2020}). \cite{Mauerhan2023} confirm the presence of CSM through the high polarisation in the emission of ATLAS17lsn. Our late-time UV detection of this source adds to the evidence for a strong interaction between the CSM and SN ejecta in this source, which was also reported by \cite{Moran2023}. 

For this source we were not able to obtain an early time image in which the SN and multiple other sources, also detected in the \textit{HST} image, were detected. Therefore, we were unable to perform relative astrometry for this source. 

To check if the observed source could be due to a young stellar cluster, we use the {\sc parsec} stellar evolutionary tracks \citep[v1.2S][]{Bressan2012, Chen2014, Tang2014, Chen2015} and {\sc colibri} evolutionary tracks \citep{Marigo2013, Rosenfield2016, Pastorelli2019, Pastorelli2020} through the CMD3.7 web interface \footnote{http://stev.oapd.inaf.it/cgi-bin/cmd} to simulate populations of 10$^4$ and 10$^5$ M$_\odot$ in mass of ages between $5\times10^5$ and $20\times10^6$~ years with steps of $5\times10^5$ years for two metallicities, [M/H]=0 (Z=0.0147) and [M/H]=$-$0.5 (Z=0.0048). The web interface has the option to output magnitudes in many filters from instruments on currently operating satellites and telescopes, including \textit{HST} WFC3/F275W. We calculate the total absolute AB magnitude of all stars in a cluster of a certain age, mass and metallicity to obtain the cluster magnitude. For the absolute magnitude we measure for the detected point source ($-$12.9~mag), there is no cluster of 10$^4$ M$_\odot$ that has an absolute magnitude brighter than the detected point source for either metallicities. For clusters with a mass of 10$^5$~M$_\odot$, for [M/H]=0 those with ages between $2\times10^5$ and $5\times10^5$ years do reach an absolute magnitude brighter than the detected point source and for [M/H]=$-$0.5 those with ages between $2.5\times10^5$ and $5\times10^5$ years do. This means there are clusters bright enough in the WFC3/F275W filter that would be detected at the distance of ATLAS17lsn. However, any underlying cluster would have to be young and unusually heavy to explain our detection. Given the previous identification of this SN as interacting, we feel that a SN is a better explanation for this detection.

\subsubsection{SN2017ivv/ASASSN-17qp}\label{sec:17qp}

ASASSN-17qp is known as a long-term emitter with different light curve slopes between 100-350~days and after 450~days (untill at least the end of the observations at $\sim700$~days; \citealp{Gutierrez2020}), of which the former is steeper than expected from the decay of $^{56}$Co and the latter is slower than expected of a source powered by the decay of $^{56}$Co. This slower decay can be explained by the presence of an additional power source, such as interaction with the CSM \citep{Gutierrez2020}, which would be consistent with our detection of this source in the UV. 

In the left panel in Figure~\ref{fig:poss_det} we can see the 3$\sigma$ uncertainty region of the position of ASASSN-17qp does not overlap with the centroid of the potential SN detection. As mentioned before, we do not include the centroiding error of the original SN detection in the calculation of the uncertainty on the SN position. Since the error on the position of this SN is 0.028\arcsec, which is one of the smallest in the sample, it is conceivable the centroiding error is not negligible for this source. For this SN, 3$\sigma = 0.084$\arcsec, while the distance between the reported SN position and the centroid position of the point source is $\sim0.09$\arcsec, which means there is only a 0.006\arcsec difference. Assuming the FWHM$\approx1$\arcsec\, during the discovery observation (or on the follow-up observations done for the ASAS-SN sources), $\sigma_{centroid}>0.006$\arcsec~if in the discovery image the detection SNR<70. It is conceivable the SNR was lower than 70, meaning the centroiding error was non-negligible when calculating the uncertainty in the position of this particular SN. We therefore consider this point source to be associated with the SN.

\begin{figure}
 \centering
    \includegraphics[width=.4\textwidth]{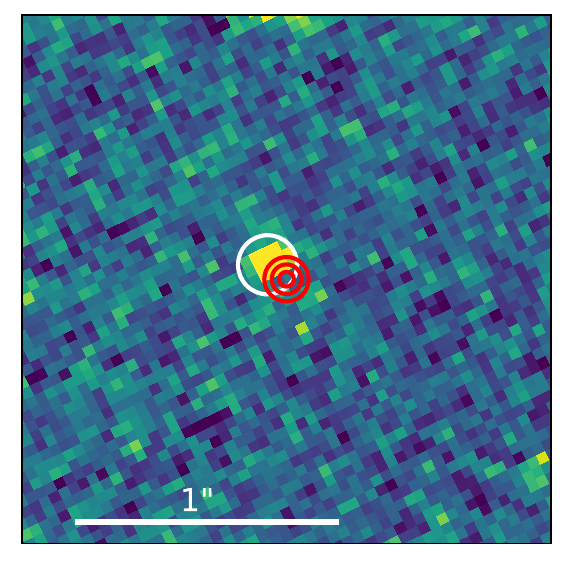}
    \caption{Same as the left panel of Figure~\ref{fig:poss_det}, except the white circle represents the projected position of a detection of ASASSN-17qp from a ground based image using relative astrometry. The size of the white circle is the 3$\sigma$ uncertainty on the projected position, defined as the RMS of the solution of the mapping of the relative pixel positions of the stars in the ground based image to those of the \textit{HST} image. North is up and East is left.}
 \label{fig:rel_ast}
\end{figure}

We obtained an (acquisition) image from the ESO archive. This image was taken on 2018 March 25, 103 days after the discovery, with the EFOSC2 instrument on the NTT. The SN is still detected on this image. We use {\sc iraf} to perform relative astrometry between this image and our \textit{HST} images using the {\sc geomap} and {\sc geoxytran} tasks to map the SN position on the ground based image to that on the \textit{HST} image. Figure~\ref{fig:rel_ast} shows the \textit{HST} image, centred on the position of ASASSN-17qp, with the position of the SN obtained from the ground indicated by a white circle. The size of the white circle represents the 3$\sigma$ RMS uncertainty of the {\sc geomap} transformation between the ground based image and our {\sc HST} images, which is 2.8 pixels (0.11~arcsecs). The red circles show the 1, 2 and 3$\sigma$ uncertainty region of the reported SN position. From Figure~\ref{fig:rel_ast} we can clearly see that our detected source lies within the 3$\sigma$ uncertainty region of the SN position as determined from the ground based image. The uncertainty region also overlaps with the 3$\sigma$ uncertainty region of the full SN position as reported in Table~\ref{tab:sample}. We therefore conclude our detection is indeed most likely still the SN. The chance of a false positive due to the SN position coinciding with pre-existing emission out to the offset of the SN is ${3\times10^{-4}}$. 

For the simulated cluster absolute magnitudes as described above, all clusters in the age, mass and metallicity range reach an absolute magnitude brighter than the absolute magnitude of the detected point source ($-$6.7~mag). However, due to the proximity of ASASSN-17qp and its host galaxy, we would expect to see resolved clusters and not point sources. In this work we therefore assume this point source is associated with the SN. 

\subsection{Comparison to known interacting SNe}

Our comparison of the detections of the interacting SNe in this work with those of well known interacting SNe is hampered by the fact that the combination of late-time and UV imaging observations as presented here is unique. For example, SN2010jl has been observed for well over 1500 days, which is a similar timescale to our sample, but those observations were only done in optical filters. \cite{Fransson2014} report one measurement in the \textit{u'}-band (their bluest band and close enough to the F275W filter to allow comparison) at 917~days of 19.16$\pm$0.12~mag, which converts to an absolute magnitude of -14.28$\pm$0.12 using the distance measure from \cite{Jencson2016}. If we compare that to the upper limits plotted in Figure~\ref{fig:UL_lightcurve} we see that this particular SN would have been detected in all but one of the images, even on top of extended emission and even if the source would be slightly dimmer in the F275W filter. Another example, iPTF14hls, was monitored for more than 1200~days in optical by \cite{Sollerman2019}, however, after $\sim$950~days there are no more observations in the bluest band in their data. SN2013L was also monitored up to 1500~days, but only in the \textit{ir}-bands \citep{Andrews2017}.

There are, however, late-time UV spectroscopic measurements of SN2010jl, taken with the STIS instrument on board \textit{HST}. The G230L(B) gratings cover the wavelength range of the WFC3/F275W filter used in this work. We obtain two spectra taken at 1224 and 1618~days and use the {\sc Python} package {\sc pyphot} (v1.4.7; \citealt{pyphot}) to perform synthetic photometry on these spectra. The obtained absolute magnitudes (corrected for Galactic extinction) are plotted in Figure~\ref{fig:UL_lightcurve}, from which we can see that in the majority of images a SN similar to SN2010jl would have been detected. For another SN, SN1993J, there are late time UV-spectra obtained with the FOS instrument aboard \textit{HST} \citep{Fransson2005}. We follow the same procedure to obtain synthetic photometric measurements in the WFC3/F275W filter from the spectra and added these measurements to Figure~\ref{fig:UL_lightcurve} as well. In the majority of the images, a SN similar to SN1993J would be below the detection limit and would therefore be undetected. SN1993J is less bright than SN2010jl and ATLAS17lsn at similar times, but brighter than ASASSN-17qp. This again supports the argument that there is a wide variety in the properties of interacting SNe and that more monitoring at late times is needed to fully understand the spectral and photometric characteristics of interacting SNe.

\subsection{Comparison to interacting SN models}

Comparing our work to models of CSM interacting SNe from e.g., \cite{Dessart2023} is complicated without extensive modelling. However, we have over plotted the light curve from \cite{Dessart2023} in the \textit{UVW2}-filter (which is similar to the F275W filter used in this work) presented in their figure~3 in our light curve in Figure~\ref{fig:UL_lightcurve}. The specific model that is discussed in detail in \cite{Dessart2023} is a 15.2~M$_\odot$ star that evolves into a red super giant (RSG) and the resulting SN is injected with an extra power of 10$^{40}$~erg~s$^{-1}$, corresponding with pre-SN wind of $\sim10^{-6}$~M$_\odot$~yr$^{-1}$. For details on the other parameters such as mixing and the numerical setup, see \cite{Dessart2023} and references therein. \cite{Dessart2023} argue the amount of mass loss in this model is a typical value for a RSG and therefore the results are representative expectations for a standard Type II SN. Lower values of injected energy would imply a negligible impact of wind mass loss on the progenitor evolution. For different values of the injected power the effect on the brightness evolution is qualitatively similar. In respect to our work this means that more energy injection will mean a brighter UV source at later times, but is unclear from their work how much brighter, so we limit our self to comparing to their model discussed in detail. It does mean that the model we compare to can be considered a lower limit to the brightness for a star of this mass.

If we compare the model of \cite{Dessart2023} represented by the black line until 1000~days and extrapolated with the same decay rate to later times in Figure~\ref{fig:UL_lightcurve} to the upper limits and detections we obtained, we can see that in roughly half of the images a SN as bright as this model would have been detected at late times. This means that for those images a non-detection does constrain the wind mass loss rate to being negligible, maybe even going as far as telling us that there was no CSM at the time of the SN. For the other upper limits a SN represented by this particular model would not have been detected as it is fainter than the upper limit of these images. We note that there are three images where this model would have been detected even on top of an extended emission region. SN2010jl is brighter than this model, but has a steeper decay rate between the two data points. SN1993C follows the same decay rate as the model, but is less bright. As we just argued that this particular model form \cite{Dessart2023} can be considered a lower limit on the luminosity, it is clear this model also does not represent SN2010jl in particular.

\subsection{Implications for peculiar transient AT2018cow}

Figure~\ref{fig:UL_lightcurve} enables us to compare the UV light curve of the CCSNe in our sample to the UV light curve of the peculiar transient \cow\ and the interacting SN model from \cite{Dessart2023}. \cite{Sun2022} were the first to report the presence of a UV bright source at the position of AT2018cow at late times (>2 years), which they use to disfavour the origin of AT2018cow as a tidal disruption event (TDE). Comparing the magnitudes of the two possible SN detections in our sample to the light curve of AT2018cow, there is no evidence that the light curve of AT2018cow is atypical for interacting SNe. Comparing to the model from \cite{Dessart2023} mentioned before, we see that \cow\ seems to decay slower than this model, but for the last epoch the magnitude is similar. However, there are many parameters in the models from \cite{Dessart2023} that can be changed (initial stellar mass, mixing parameters, shape of the CSM, etc.). Those parameters can influence the brightness and decay rate, so we cannot draw any firm conclusions from this comparison. 31 out of 49 of the SNe for which we do not detect a point source have a limiting magnitude brighter than AT2018cow, meaning a source as bright as \cow\ can be present in these images without it being detected and the remaining 18 have a limiting magnitude fainter than AT2018cow at their respective observational epochs. Among the possible SN detections, one has a larger distance than AT2018cow (ATLAS17lsn), and one has a smaller distance (ASASSN-17qp). ATLAS17lsn has an absolute magnitude $\sim3$ magnitudes brighter than AT2018cow, while ASASSN-17qp has an absolute magnitude that is $\sim3$ magnitudes fainter than AT2018cow. Combined, this means the presence of late-time UV emission in AT2018cow is not an argument against a SN nature of this event. The difference in $\Delta t$ (the time between discovery and the reported detection in this work) between the observations of \cow\ and the detected SNe are not relevant for the conclusion, given the slow light curve evolution of \cow.

If we limit the sample to SNe closer than \cow, we can investigate if sources as bright as \cow\ would have been detected if present or if emission as bright as \cow\ would fall under the detection limit of the respective image. For emission as bright as \cow\ at distances closer than \cow, we would expect the emission to be brighter than the limiting magnitudes of the images under normal circumstances and so limiting to this distance range is a good probe of whether or not \cow\ is unusually bright compared to the other CCSNe in this distance range. Only 5 out of 17 CCSNe have an absolute magnitude (limit) brighter than AT2018cow and the remaining 12 have a fainter limit (see Fig~.\ref{appfig:lightcurve} in the appendix). For this smaller sample it means that SNe as bright as \cow\ would have been detected in 75~percent of the images. The model of \cite{Dessart2023} would also have been detected in 75~percent of the images, and as we do not detect a SN in this many images, we conclude \cow\ is brighter than we would expect for CCSNe out to the distance of \cow. This is an argument against a SN nature for \cow. 

\cow\ is a black body (BB) with a very small radius ($\sim40$~R$_\odot$ at 713 and 1474~days; \citealt{Inkenhaag2023}). For any SN with CSM, if it is a BB at these epochs, then the photosphere is in the CSM, which means the radius will be bigger than 40 Rsun \citep[e.g.,][]{Fassia2001, Smith2010, Kumar2019, Wang2023, Sfaradi2024}. Therefore AT2018cow is unlikely to be a CCSN, in agreement with e.g., \cite{Liu2018}.

\section{Conclusions}

Using our sample of 51 CCSNe observations we only find two point source detections in {\it HST} data taken between two and five years after discovery. The absolute UV magnitude of the two detections compared to the late-time absolute UV magnitude of AT2018cow provides no evidence that the late-time UV detection of AT2018cow is an argument against a SN nature. AT2018cow is, however, bright when compared to the subset of our CCSNe sample that is as nearby as, or lies at a distance closer than AT2018cow. Combined with the photospheric radius of AT2018cow's emission being orders of magnitude smaller than that of CCSNe, we conclude that the late-time UV emission of AT2018cow is not likely to be driven by interaction. One explanation for its late-time UV emission is that we are viewing the inner region of the explosion, which could be a long-lived accretion disk. Such a scenario is expected in TDE models, and less likely in SN scenarios. 

These two detections represent some of the first UV detections of core collapse supernovae at late times after the explosion. Both SNe were also previously reported as interacting SNe, which means the late-time emission may be explained by continuing interaction between the SNe ejecta and the CSM that is expelled by the progenitor before the explosion \citep[e.g.][]{Dessart2022, Dessart2023}. 
While we know the percentage of interacting SNe is $\sim$10 percent \citep[e.g.,][]{Li2011, Smith2011}, obtaining multiple epochs of UV imaging, in different bands, with contemporaneous spectra and detailed modelling will be necessary to fully understand these events (see \citealt{Fransson2014} for a good example of what can be achieved). In most SNe, timescales of two to five years after explosion are therefore also suitable for investigating the SN environment.
Furthermore, archival searches to look for detections of the progenitor star will give more direct information about the type of stars that can create these interacting SNe and can give information about the evolutionary stages in which CSM creation happens.

\begin{acknowledgements}
A.I.~would like to thank Zheng Cao for helpful discussions about the sizes of the emission regions. The authors would like to thank the anonymous referee for the feedback provided, which has significantly improved the paper.

This work is part of the research programme Athena with project number 184.034.002, which is financed by the Dutch Research Council (NWO).
The scientific results reported on in this article are based on data obtained under \textit{HST} Proposal 16287 with PI J.D.~Lyman.

P.G.J.~has received funding from the European Research Council (ERC) under the European Union’s Horizon 2020 research and innovation programme (Grant agreement No.~101095973).
M.F. is supported by a Royal Society - Science Foundation Ireland University Research Fellowship.
J.D.L. acknowledges support from a UK Research and Innovation Future Leaders Fellowship (MR/T020784/1).
L.G. acknowledges financial support from the Spanish Ministerio de Ciencia e Innovaci\'on (MCIN) and the Agencia Estatal de Investigaci\'on (AEI) 10.13039/501100011033 under the PID2020-115253GA-I00 HOSTFLOWS project, from Centro Superior de Investigaciones Cient\'ificas (CSIC) under the PIE project 20215AT016 and the program Unidad de Excelencia Mar\'ia de Maeztu CEX2020-001058-M, and from the Departament de Recerca i Universitats de la Generalitat de Catalunya through the 2021-SGR-01270 grant.
H.K. was funded by the Research Council of Finland projects 324504, 328898, and 353019.

This work makes use of Python packages {\sc numpy} \citep{2020Natur.585..357H}, {\sc scipy} \citep{2020NatMe..17..261V}; {\sc matplotlib} \citep{2007CSE.....9...90H}, {\sc extinction} \citep{barbary_kyle_2016_804967} and {\sc pandas} \citep[v1.3.5;][]{mckinney2010data, pandas}
This work made use of Astropy (\url{http://www.astropy.org}): a community-developed core Python package and an ecosystem of tools and resources for astronomy \citep{astropy:2013, astropy:2018, astropy:2022}.
\end{acknowledgements}

{\scriptsize
\noindent References for Table~\ref{tab:sample}
\newline (a)\cite{15bm} (b)\cite{Holoien2017a} (c)\cite{16bmi} (d)\cite{Holoien2017b} (e)\cite{16bsb} (f)\cite{16gy} (g)\cite{16egz} (h)\cite{16gkg} (i)\cite{16hbb} (j)\cite{16hgm} (k)\cite{16hvu} (l)\cite{16ai} (m)\cite{Brennan2022} (n)\cite{16jbu} (o)\cite{17bif} (p)\cite{Holoien2019} (q)\cite{17caw} (r)\cite{17dch} (s)\cite{17dek} (t)\cite{17dhu} (u)\cite{17dka} (v)\cite{17dkb} (w)\cite{17eca} (x)\cite{17hpc} (y)\cite{17fbu} (z)\cite{17ffm} (aa)\cite{17ffq} (ab)\cite{17ffm} (ac)\cite{17fod} (ad)\cite{17fqk} (ae)\cite{17fqo} (af)\cite{17fvr} (ag)\cite{17fwm} (ah)\cite{17ggw} (ai)\cite{17gip} (aj)\cite{17giq} (ak)\cite{17chn} (al)\cite{17lsn} (am)\cite{17oj} (an)\cite{17ipa} (ao)\cite{17ivu} (ap)\cite{Gutierrez2020} (aq)\cite{17qp} (ar)\cite{17jbj} (as)\cite{17mw} (at)\cite{18eo} (au)\cite{Neumann2023} (av)\cite{18ant} (aw)\cite{18anu} (ax)\cite{18dfg} (ay)\cite{18eog} (az)\cite{18evy} (ba)\cite{18fit} (bb)\cite{18iuq} (bc)\cite{18jfz} (bd)\cite{18cb}}

\bibliographystyle{aa}
\bibliography{references}

\onecolumn 
\begin{appendix}

\section{Supernova positions and uncertainties}

 \begin{figure*}[h!]
 \centering
    \includegraphics[width=.84\textwidth]{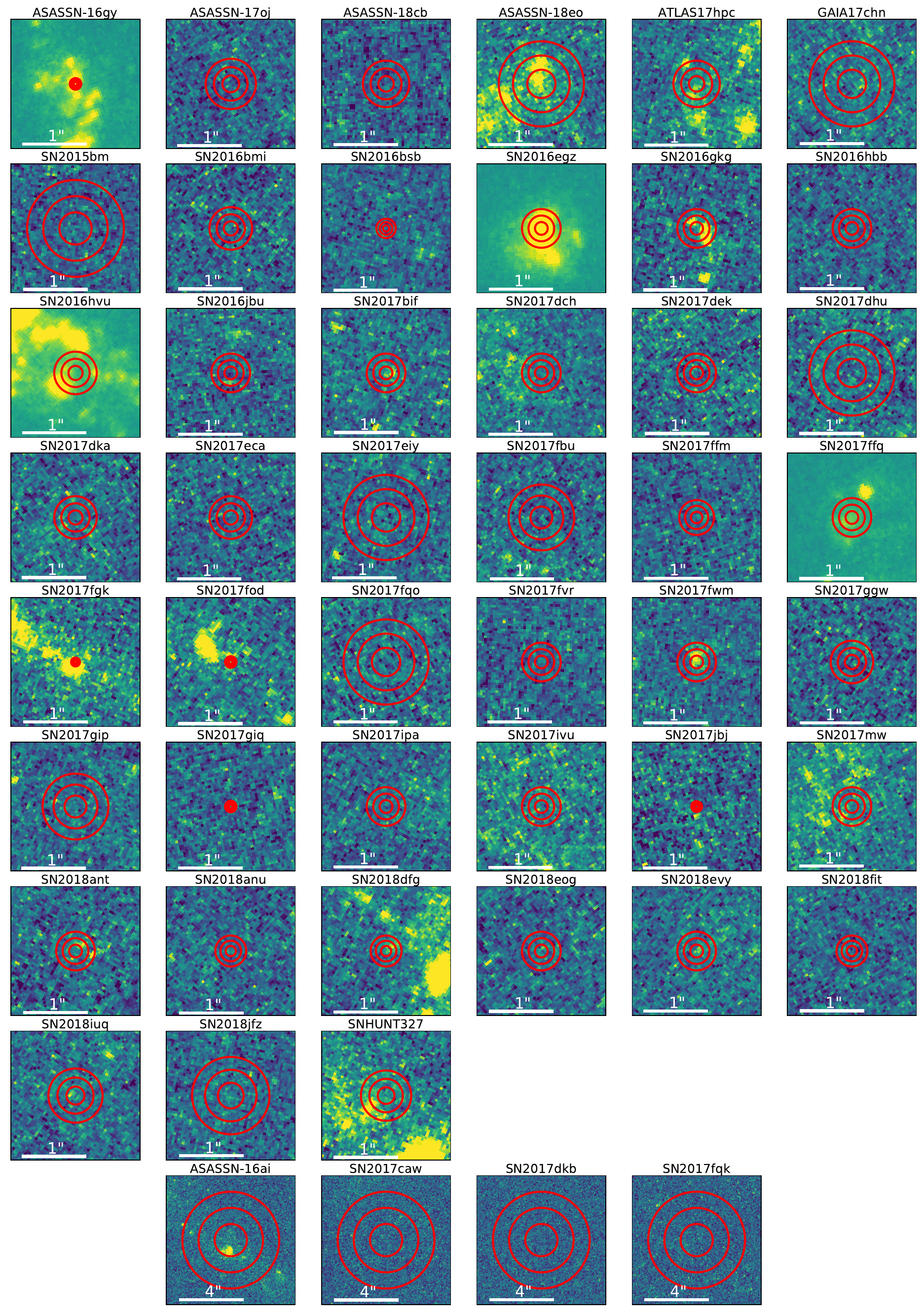}
    \caption{Mosaic showing different size
    %$2\arcsec \times 2\arcsec$ 
    cutouts of all SNe in our sample where we did not detect a point source within the 3$\sigma$ positional uncertainty region. The cutouts are centred on the SN positions. The 1, 2 and 3$\sigma$ errors on the position is shown by the red circles. The white lines indicate a length of 1\arcsec (or 4\arcsec if the 1$\sigma$ uncertainty region is 1\arcsec, separated on the bottom row). For all images North is up and East is left. The scale of the pixel values was chosen to show any structure present in the emission.}
 \label{fig:SN_mosaic}
\end{figure*}

\newpage

\section{SN2018ant individual frames}

 \begin{figure*}[h!]
 \centering
    \includegraphics[width=.84\textwidth]{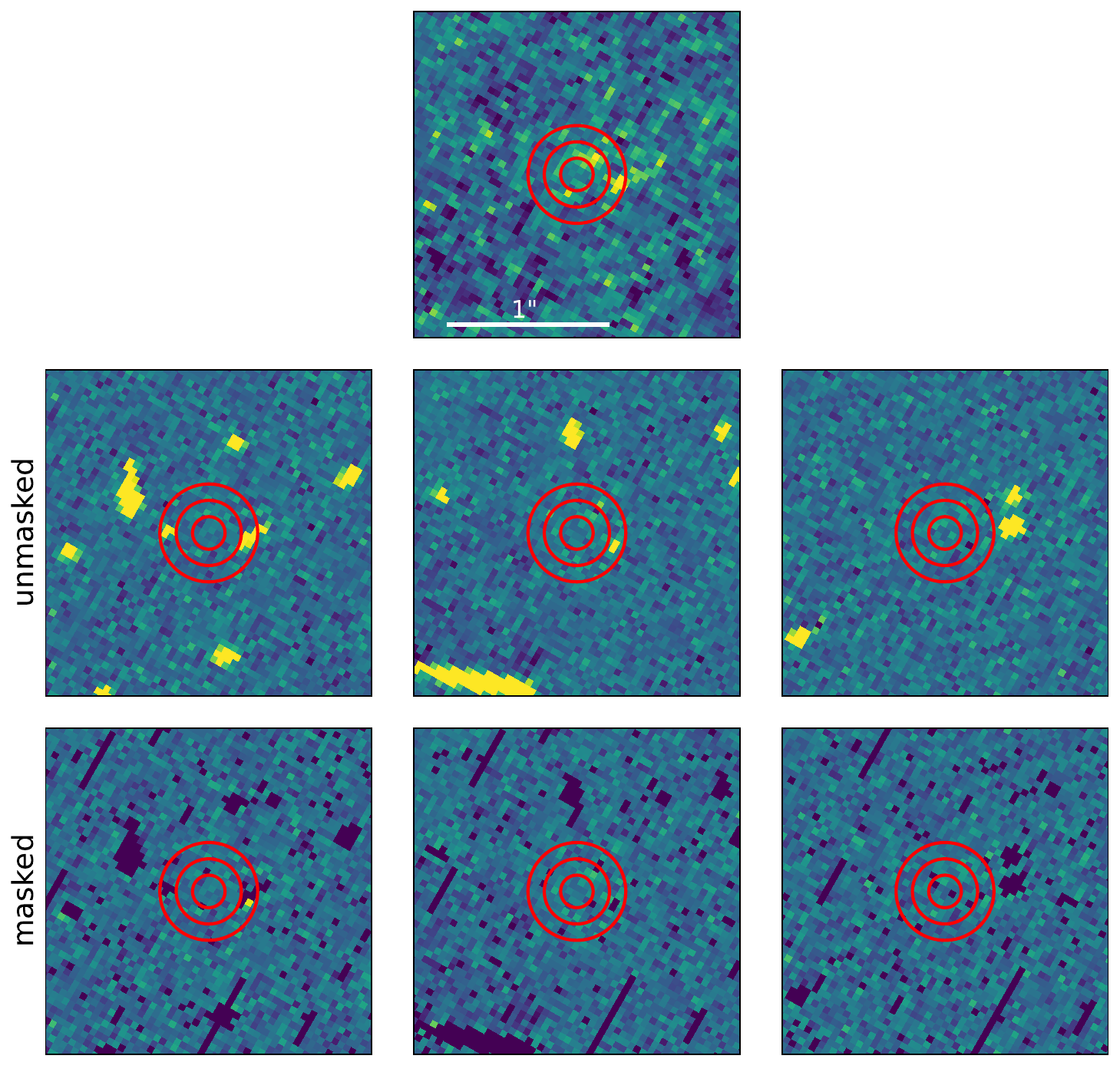}
    \caption{Mosaic showing the $2\arcsec \times 2\arcsec$ cutout of the position of SN2018ant in the top row, its individual \texttt{\_flc} images in the middle row and the same individual \texttt{\_flc} images after masking by {\sc DOLPHOT} in the bottom row. The cutouts are centred on the SN position. The 1, 2 and 3$\sigma$ error on the source position is indicated by the red circles in each image. The white line indicates a length of 1\arcsec. For all images North is up and East is left. The scale of the pixel values was arbitrarily chosen.}
 \label{appfig:18ant_flc}
\end{figure*}

\newpage

\section{Sources closer than \cow}

 \begin{figure*}[h!]
 \centering
    \includegraphics[width=.84\textwidth]{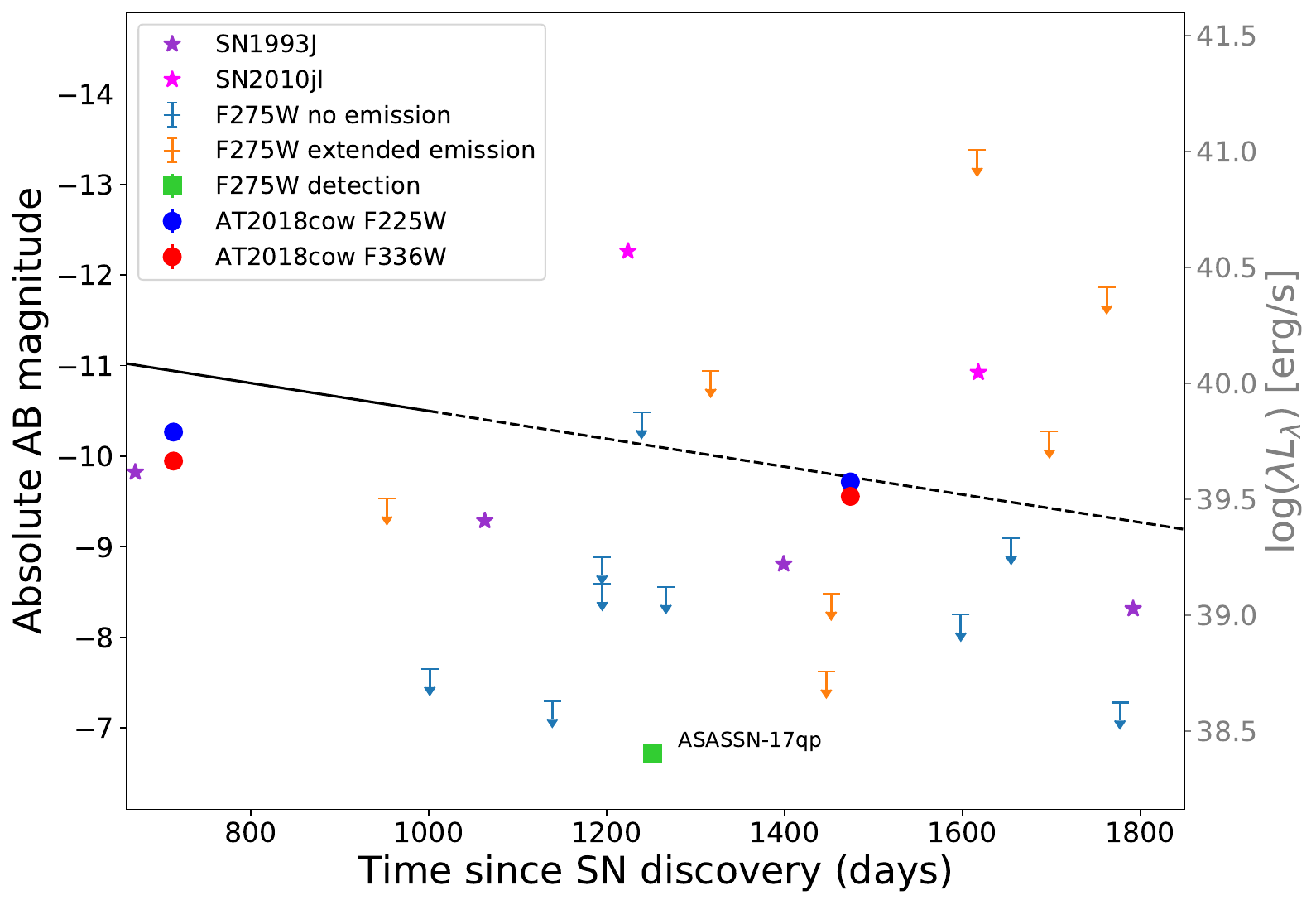}
    \caption{Same as Figure~\ref{fig:UL_lightcurve}, but showing only the sources with a distance closer than the distance to \cow.}
 \label{appfig:lightcurve}
\end{figure*}

\end{appendix}

\end{document}